# Sensor-Outage-Aware Spatio-Temporal Graph Reconstruction of High-Rise Façade Pressure Fields

**Seyedeh Fatemeh Mirfakhar[1], Reda Snaiki[1*]**

[1] *Department of Construction Engineering, École de Technologie Supérieure, Université du Québec, Montréal, Québec, Canada*
[*] *Corresponding author*. *Email*: reda.snaiki@etsmtl.ca

**Abstract:** Time-resolved façade pressure fields are essential for the wind-resistant design and aerodynamic assessment of high-rise buildings. However, dense instrumentation is costly and often impractical, and sensor outages can further reduce data availability. This study proposes a sensor-outage-aware spatio-temporal graph reconstruction framework for completing façade pressure fields from sparse measurements. The method couples temporal feature extraction with graph-based spatial propagation on a unified façade-domain representation and uses an explicit observation-availability indicator to handle temporarily unavailable sensor signals while reconstructing both missing instrumented channels and non-instrumented locations. The framework is evaluated using wind-tunnel pressure coefficient data for a high-rise building across windward, lateral, and leeward façades under multiple wind directions. The results show reliable outage-tolerant reconstruction at instrumented sensors and accurate full-field completion at non-instrumented nodes, with reconstruction generally most accurate on the windward façade and more challenging on the lateral and leeward façades. Time-domain, spectral, and spatial validations further show that the framework preserves the dominant temporal evolution, principal dynamic content, and coherent large-scale pressure-field organization, while the largest residual discrepancies remain localized in higher-frequency or intermittent components. A two-stage predictive extension is also outlined, in which future sensor signals are forecast at available instrumented locations and then mapped to future full-field pressure estimates through the proposed reconstruction model.

## 1. Introduction

High-rise buildings are particularly susceptible to wind loading because their height, slenderness, and lower effective damping can amplify dynamic response and serviceability demands [1–2]. Wind effects on tall buildings extend beyond global actions such as base shear and overturning moments to include façade cladding demands, local peak suctions, and spatially varying pressure patterns that influence envelope performance and occupant comfort [3–6]. Accurate assessment of these effects requires time-resolved façade pressure information, since both mean and fluctuating components contribute to design-relevant quantities such as peak effects, spatial correlation, and frequency-dependent response [7]. From an aerodynamic standpoint, façade pressures are governed by flow mechanisms such as impingement, separation, shear-layer development, and wake effects, which vary across windward, lateral, and leeward surfaces and depend on wind direction and upstream turbulence [8–10]. These mechanisms generate pressure fields with strong spatial gradients near edges and corners and more irregular temporal behavior on side and leeward façades, where separated flow and wake-induced fluctuations can be broadband and intermittent [11–13]. Capturing this complexity is essential for estimating local cladding demands and reproducing the spatial coherence required for system-level response and reliability assessment. Consequently, obtaining dense, time-resolved façade pressure fields remains a central challenge in wind engineering for high-rise buildings [14–16].

Façade pressure fields can be obtained through computational, experimental, and full-scale measurement approaches, each with distinct strengths but also important practical limitations. CFD provides flexible access to surface-pressure distributions and parametric studies, but high-fidelity prediction around bluff high-rise buildings remains computationally demanding and sensitive to turbulence modeling, inflow specification, and spatial resolution, particularly in separated-flow regions [9, 17–19]. Wind-tunnel testing remains a standard tool in building aerodynamics and can provide detailed façade-pressure measurements under controlled boundary-layer conditions, but dense pressure-tap layouts are costly and experimentally demanding to implement [10, 20–21]. Full-scale measurements offer the most realistic operating conditions, yet they are usually restricted to sparse sensor networks and are further affected by noise, environmental nonstationary, and missing data caused by sensor malfunction or communication loss [22–23]. Consequently, dense time-resolved façade pressure fields are rarely available in practice, and pressure monitoring and experimental studies often rely on sparse instrumentation. This limitation has motivated substantial research on sensor placement strategies under limited sensing budgets [24–26]. However, even an optimized sensor layout does not eliminate the need for field reconstruction, since many engineering quantities of interest, such as full-field pressure maps, spatial correlation, and pressure patterns associated with separation and wake effects, require spatially dense information [11, 26]. A further practical complication is that sensor availability cannot be assumed to remain fixed: instrumented channels may become temporarily unavailable because of malfunction, maintenance, drift, or communication loss [27–29]. These challenges highlight the need for reconstruction strategies capable of recovering dense pressure fields from sparse and imperfect measurements.

A range of approaches has been explored to infer dense façade pressure information from limited measurements, including spatial interpolation and regression strategies, reduced-order reconstructions based on low-dimensional pressure representations, and more recent machine-learning models that learn nonlinear mappings from sparse observations to dense fields [11, 26, 30–32]. These methods have improved reconstruction accuracy in several settings, particularly when dense training data are available and the sensing configuration is fixed [33–35]. However, important limitations remain for high-rise façade pressure reconstruction. Many methods are formulated as snapshot-based spatial inference and therefore underuse temporal coherence, which is informative for reconstructing broadband and intermittent fluctuations typical of side and leeward façades [34–37]. Methods that incorporate temporal modeling are often evaluated on limited aerodynamic regimes or a single façade type, without systematic assessment across windward, lateral, and leeward surfaces under multiple wind directions [38]. In addition, sensor locations are commonly assumed to remain fixed and continuously available, whereas practical monitoring systems must tolerate temporary sensor outages that may occur at critical locations [39–41]. These limitations motivate reconstruction frameworks that jointly address sparse sensing, missing observations, and full-field completion while exploiting both spatial structure and temporal dynamics.

To address these limitations, this study develops a sensor-outage-aware spatio-temporal graph reconstruction framework for completing high-rise façade pressure fields from sparse and imperfect measurements. Each façade is represented as a unified spatial graph, and the reconstruction model couples temporal feature extraction with graph-based spatial propagation to recover the full time-resolved pressure field over the same interval as the available sensor data. Sensor outage is treated as an inherent aspect of the problem through structured masking of instrumented channels during training, enabling the model to reconstruct both temporarily unavailable instrumented signals and non-instrumented locations. The framework is evaluated using wind-tunnel pressure coefficient data for a high-rise building across multiple wind directions and three façade regimes (i.e., windward, lateral, and leeward) allowing systematic assessment under aerodynamic conditions of increasing complexity. A two-stage predictive extension is also outlined, in which forecasted sensor signals are mapped to future full-field pressure estimates through the proposed reconstruction model.

## 2. Methodology

### 2.1 Problem definition and reconstruction objective

This study addresses the reconstruction of façade wind-pressure fields from a limited set of measurements collected at instrumented locations under partial observation. The objective is to recover the pressure response over the full façade domain, including both non-instrumented locations and temporarily unavailable instrumented sensors, using the measurements available from the remaining sensors over the same time interval. The problem is therefore formulated as a spatio-temporal reconstruction problem under constrained sensing.

Let $\mathcal{G} = (\mathcal{V}, \mathcal{E})$ denote a graph representation of the façade domain, where $\mathcal{V}$ is the set of spatial locations on the surface and $\mathcal{E}$ encodes spatial connectivity relationships. For a reconstruction interval of length $T$, the target pressure-coefficient field is written as $\mathbf{Y} \in \mathbb{R}^{|\mathcal{V}| \times T}$, where each row corresponds to a spatial location and each column corresponds to a time instant within the interval. Only a subset of locations $\mathcal{V}_s \subset \mathcal{V}$ is instrumented. In a given reconstruction scenario, some of these instrumented measurements may be unavailable due to training-time masking or simulated sensor-outage conditions. If $\mathcal{M} \subset \mathcal{V}_s$ denotes the set of temporarily unavailable instrumented nodes, then the reconstruction task is to estimate the full field from the resulting partial observations. The reconstruction mapping is written as:

$$\widehat{\mathbf{Y}} = f_\theta(\mathbf{X}, \mathbf{P}, \mathbf{C}) \quad (1)$$

where $\mathbf{X}$ denotes the partial spatio-temporal measurements, $\mathbf{P}$ is the observation mask that distinguishes available and missing entries, $\mathbf{C}$ denotes contextual conditioning variables, and $f_\theta$ is a trainable mapping parameterized by $\theta$. The output $\widehat{\mathbf{Y}}$ is the reconstructed pressure field over the full spatial domain $\mathcal{V}$ and over the same temporal interval as the input.

Under this formulation, the methodology addresses two coupled objectives. The first is the recovery of signals at temporarily unavailable instrumented nodes, which assesses the robustness of the framework to sensor loss and its ability to exploit redundancy within the sensing network. The second is the reconstruction of the response at non-instrumented locations, which constitutes the primary field-completion task. Although the inference target is spatial, accurate reconstruction also requires preservation of the time-resolved characteristics of the pressure signals and the temporal dependencies contained in the available measurements. The problem is therefore treated as a unified spatio-temporal reconstruction task under sparse and imperfect sensing.

## 2.2 Overview of the proposed framework

The proposed methodology is formulated as a unified spatio-temporal reconstruction framework that maps partially observed pressure measurements to a time-resolved reconstruction of the full façade pressure field. The target is field completion over a given interval using the measurements available over that same interval. To achieve this, the framework combines temporal representation learning, graph-based spatial information propagation, and explicit handling of observation missingness within a single end-to-end trainable architecture. Figure 1 summarizes the overall workflow.

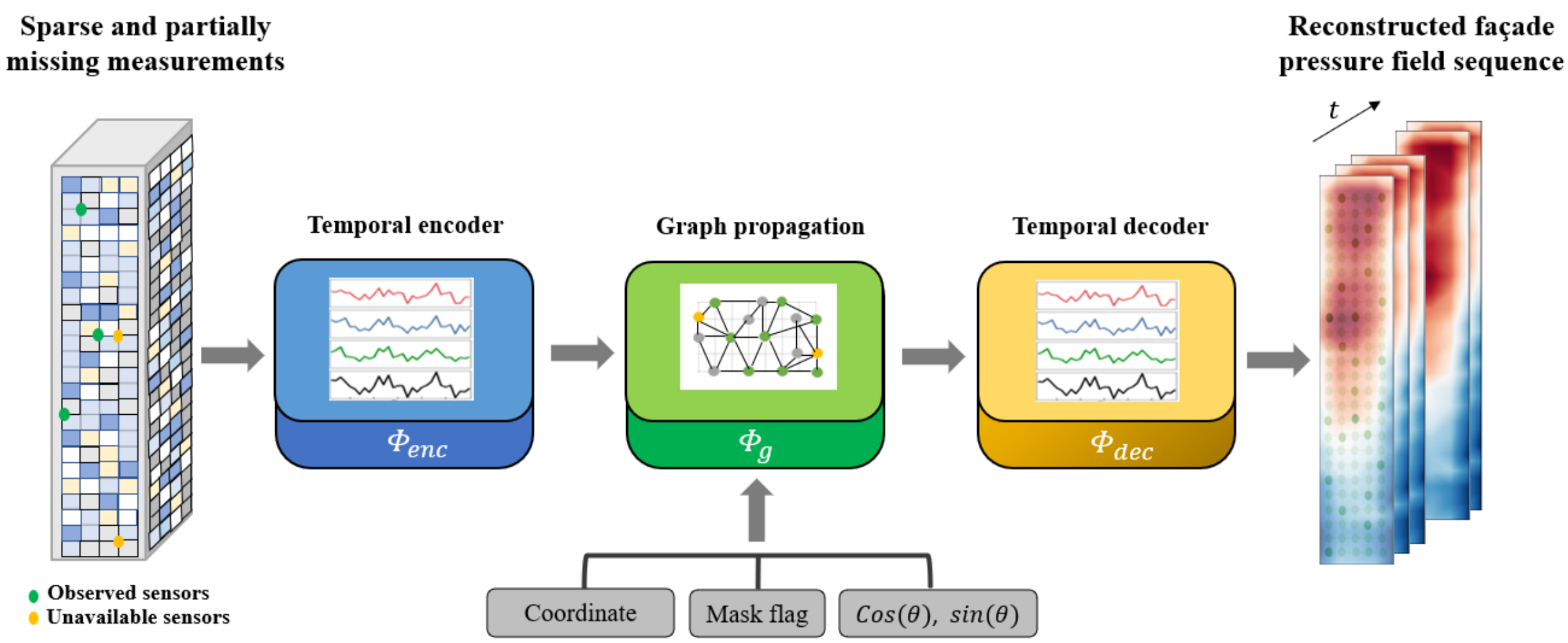

**Fig. 1** Schematic of the proposed spatio-temporal graph reconstruction framework for façade pressure field completion from sparse and partially missing measurements.

For each reconstruction interval, the input consists of a partially observed pressure field defined on the façade domain together with an observation-availability indicator that distinguishes measured and missing entries. The available signals are first transformed by a temporal encoder into latent features that summarize the local dynamic behavior over the input interval. These latent features are then processed by a graph-based propagation module defined on a unified graph representation of the façade domain, allowing information from observed locations to be transferred to unobserved regions through spatial connectivity while incorporating node-level contextual descriptors such as spatial position and measurement availability. A temporal decoder finally maps the propagated latent representation back to the physical signal space, yielding a reconstructed pressure field over the full spatial domain and the same reconstruction interval.

A key feature of the framework is that instrumented locations, temporarily unavailable sensors, and non-instrumented locations are all treated within a common graph domain rather than through separate sequential prediction stages. Reconstruction is therefore posed directly at the field level, which promotes consistency across the different reconstruction targets and avoids discontinuities that may arise when local sensor recovery and spatial interpolation are handled independently. To improve robustness under varying missing-observation patterns, temporary sensor unavailability is simulated during training by masking subsets of instrumented measurements, while the observation-availability indicator is retained as an explicit input so that the model can distinguish genuinely observed values from missing entries.

### 2.3 Spatio-temporal reconstruction network

The proposed reconstruction network is formulated as an encoder–propagator–decoder architecture operating on partially observed, time-resolved graph signals. Its purpose is to transform sparse pressure measurements into a latent representation that preserves the temporal structure of the observed signals, propagate this information across the façade domain through graph-based interactions, and reconstruct the pressure response over the full set of spatial locations for the same reconstruction interval.

For a given reconstruction interval, the model receives a partially observed signal tensor defined over the façade graph together with an observation-availability indicator. Let $N = |\mathcal{V}|$ denote the number of spatial locations in the graph and $T$ the length of the reconstruction interval. The input can be written in generic form as $\mathbf{X} \in \mathbb{R}^{N\times T}$, where unavailable entries correspond either to non-instrumented locations or to temporarily unavailable measurements from instrumented sensors. An observation mask $\mathbf{P} \in \{0,1\}^{N\times T}$ is provided in parallel to identify measurement availability. In addition, node-wise geometric descriptors and loading-condition descriptors are incorporated as contextual information so that the network can condition its reconstruction on spatial location and operating scenario. This conditioning is particularly important in façade-pressure reconstruction, where both the spatial distribution and temporal character of the response depend strongly on local position and aerodynamic loading configuration.

The first component of the architecture is a temporal encoder that acts on the time series associated with each node. Its role is to map raw or normalized pressure signals into a compact latent representation that captures short-term dynamics, oscillatory content, and transient events while suppressing noise and redundant variability. The encoder is applied in a shared manner across nodes, which enforces a common temporal feature space and improves statistical efficiency under sparse sensing. Conceptually, this stage can be written as:

$$\mathbf{Z} = \Phi_{enc}(\mathbf{X}) \tag{2}$$

where $\mathbf{Z} \in \mathbb{R}^{N\times d\times L}$ is a latent spatio-temporal representation, $d$ is the latent channel dimension, and $L$ is the latent temporal length after temporal compression. By encoding each observed signal over a finite interval rather than treating time instants independently, the model retains temporal coherence that can later support spatial inference at unobserved locations.

The latent representation is then processed by a graph-based spatial propagation module defined on the unified façade graph. Instrumented nodes, temporarily unavailable sensors, and non-instrumented locations are all represented within the same graph, and reconstruction is therefore performed directly at the field level rather than through separate local recovery and interpolation stages. In the implemented framework, spatial propagation is performed with an attention-based message-passing operator, allowing the influence of neighboring nodes to be learned adaptively from the latent features and contextual descriptors. At each latent time slice, the node embedding is augmented with spatial and observational context, including node coordinates, availability indicator, and loading-condition encoding, and graph message passing is applied to propagate information from observed regions toward unobserved nodes. This yields the updated latent representation:

$$\tilde{\mathbf{Z}} = \Phi_g(\mathbf{Z}, \mathcal{G}, \mathbf{P}, \mathbf{C}) \tag{3}$$

where $\Phi_g$ denotes the graph propagation operator and $\mathbf{C}$ collects contextual conditioning variables. The explicit use of the observation-availability indicator at this stage is important because, in pressure reconstruction, a missing measurement cannot be reliably distinguished from a physically meaningful low-amplitude pressure value based on magnitude alone. Providing mask information to the network therefore reduces ambiguity in the learned reconstruction mapping and improves robustness when the missing-observation pattern changes between training and evaluation.

Following graph propagation, a temporal decoder maps the enriched latent representation back to the pressure signal space over the full set of nodes. The decoder operates on the propagated latent tensor and reconstructs the time-resolved response for the same interval used at the input:

$$\hat{\mathbf{Y}} = \Phi_{dec}\left(\tilde{\mathbf{Z}}\right) \tag{4}$$

with $\hat{\mathbf{Y}} \in \mathbb{R}^{N \times T}$. The decoding stage is formulated as a window-level reconstruction rather than a pointwise regression at isolated time instants. This design is consistent with the intended use of temporal information in the framework: temporal structure is exploited to improve reconstruction fidelity and stability, even though the task remains reconstruction over a known interval rather than forecasting beyond that interval. Combining the three components, the overall network may be written as:

$$\hat{\mathbf{Y}} = \Phi_{dec}\left(\Phi_g(\Phi_{enc}(\mathbf{X}), \mathcal{G}, \mathbf{P}, \mathbf{C})\right) \tag{5}$$

This composition highlights the respective roles of the temporal encoder, spatial propagation module, and temporal decoder while making clear that the reconstruction is conditioned simultaneously on graph structure, observation availability, and loading context. The formulation is general enough to accommodate different architectural instantiations of each block; the specific implementation choices adopted in this study are reported in the case-study section.

### 2.4 Training strategy and objective function

The training procedure is designed to make the reconstruction network robust to sparse instrumentation and temporary sensor unavailability while preserving fidelity in the reconstructed pressure signals over the full domain. To this end, the learning strategy combines structured masking during training with a multi-component objective function that enforces reconstruction fidelity in complementary signal domains and regularizes the latent representation under partial observation. The resulting formulation is intended to improve not only pointwise accuracy, but also temporal consistency and robustness under varying missing-observation patterns.

A central element of the training strategy is the simulation of temporary sensor loss during learning. For each training sample, the model receives a partially observed spatio-temporal field in which non-instrumented locations are unobserved by definition, while an additional subset of instrumented measurements is intentionally masked. This masking is applied over the full reconstruction interval, so that selected instrumented channels are treated as unavailable throughout the input segment even though their target signals remain available for supervision. By varying the masked subset during training, the network is encouraged to learn distributed spatio-temporal dependencies rather than relying on a fixed sensor configuration or memorized local correlations.

Let $\mathcal{V}_m$, $\mathcal{V}_o$, $\mathcal{V}_u$ denote, respectively, the sets of temporarily masked instrumented nodes, available instrumented nodes, and non-instrumented nodes in a given training sample. These three groups play different roles in the reconstruction task and are therefore supervised separately. The masked instrumented nodes provide a direct measure of sensor-recovery capability under controlled missingness, the available instrumented nodes preserve consistency with observed signals, and the non-instrumented nodes represent the primary field-completion target. This partitioned supervision aligns the optimization objective with the dual goals defined in Section 2.1 while preventing any single node category from dominating the learning process.

For a generic subset of nodes $\mathcal{S} \subseteq \mathcal{V}$, the signal reconstruction term is defined as a composite loss of the form:

$$\mathcal{L}_{sig}(\mathcal{S}) = \mathcal{L}_{amp}(\mathcal{S}) + \lambda_d \mathcal{L}_{diff}(\mathcal{S}) + \lambda_f \mathcal{L}_{spec}(\mathcal{S}) \quad (6)$$

where $\mathcal{L}_{amp}$ penalizes amplitude mismatch in the reconstructed time histories, $\mathcal{L}_{diff}$ penalizes mismatch in temporal differences, and $\mathcal{L}_{spec}$ penalizes discrepancy in spectral content. The amplitude term promotes local signal fidelity, the temporal-difference term helps preserve transient behavior and slope changes, and the spectral term encourages consistency in oscillatory content and energy distribution across frequencies. This combination is useful in façade-pressure reconstruction because acceptable performance cannot be assessed solely by pointwise agreement if the reconstructed signals fail to preserve their dynamic character. Using this subset loss, the reconstruction objective over the three node categories is written as:

$$\mathcal{L}_{rec} = \omega_m \mathcal{L}_{sig}(\mathcal{V}_m) + \omega_o \mathcal{L}_{sig}(\mathcal{V}_o) + \omega_u \mathcal{L}_{sig}(\mathcal{V}_u) \quad (7)$$

where $\omega_m$, $\omega_o$, and $\omega_u$ are weighting coefficients controlling the relative emphasis on masked sensor recovery, consistency at available sensors, and reconstruction at non-instrumented nodes. This weighted formulation is appropriate because the three subsets differ both in size and in functional role. In particular, assigning sufficient weight to the masked instrumented nodes promotes robustness to sensor outages, while retaining a nonzero contribution from available sensors stabilizes training and preserves fidelity at observed channels.

In addition to signal-space supervision, the training objective includes a latent consistency regularization term that stabilizes the representation learned under partial observation. Let $\mathbf{Z}^{\text{full}}$ denote the latent representation obtained from a reference forward pass using the unmasked input, and let $\mathbf{Z}^{\text{mask}}$ denote the latent representation obtained from the corresponding masked-input forward pass. A generic latent consistency term can then be written as:

$$\mathcal{L}_{lat} = \mathcal{D}\left(\mathbf{Z}_{\mathcal{V}_m}^{mask}, \mathbf{Z}_{\mathcal{V}_m}^{full}\right) \tag{8}$$

where $\mathcal{D}(.,.)$ denotes a distance measure in latent space. This regularization encourages masked-input latent features to remain close to those obtained under full observation, thereby improving representation stability and reducing degradation when sensor availability changes. The complete training objective is therefore expressed as:

$$\mathcal{L} = \mathcal{L}_{rec} + \lambda_z \mathcal{L}_{lat} \tag{9}$$

where $\lambda_z$ controls the strength of latent consistency regularization. The coefficients $\lambda_d$, $\lambda_f$, and $\lambda_z$, as well as the subset weights $\omega_m$, $\omega_o$, and $\omega_u$, govern the balance between local amplitude fidelity, temporal dynamics, spectral consistency, and representation stability. Their numerical values are implementation-specific and are therefore reported later in the case-study and experimental setup section. Optimization is performed end-to-end on randomly sampled reconstruction intervals under varying masking patterns using standard gradient-based training, with the observation-availability indicator retained as part of the model input throughout.

Under this formulation, the training objective is directly aligned with the operational requirements of sparse and imperfect sensing. By combining structured masking, subset-aware supervision, multi-domain signal fidelity constraints, and latent consistency regularization, the network is trained to reconstruct time-resolved pressure fields that remain accurate at both observed and unobserved locations under changing sensor-availability conditions.

### 2.5 Inference procedure for full-sequence reconstruction

The reconstruction network introduced in Sections 2.2–2.4 operates on finite temporal intervals. In practical use, however, the objective is to recover a continuous, time-resolved pressure field over records that are typically much longer than the fixed window length used during training. The inference procedure must therefore extend the learned window-level reconstruction operator to the full sequence while preserving temporal continuity and reducing boundary artifacts associated with segmented processing. The adopted workflow is illustrated schematically in Fig. 2.

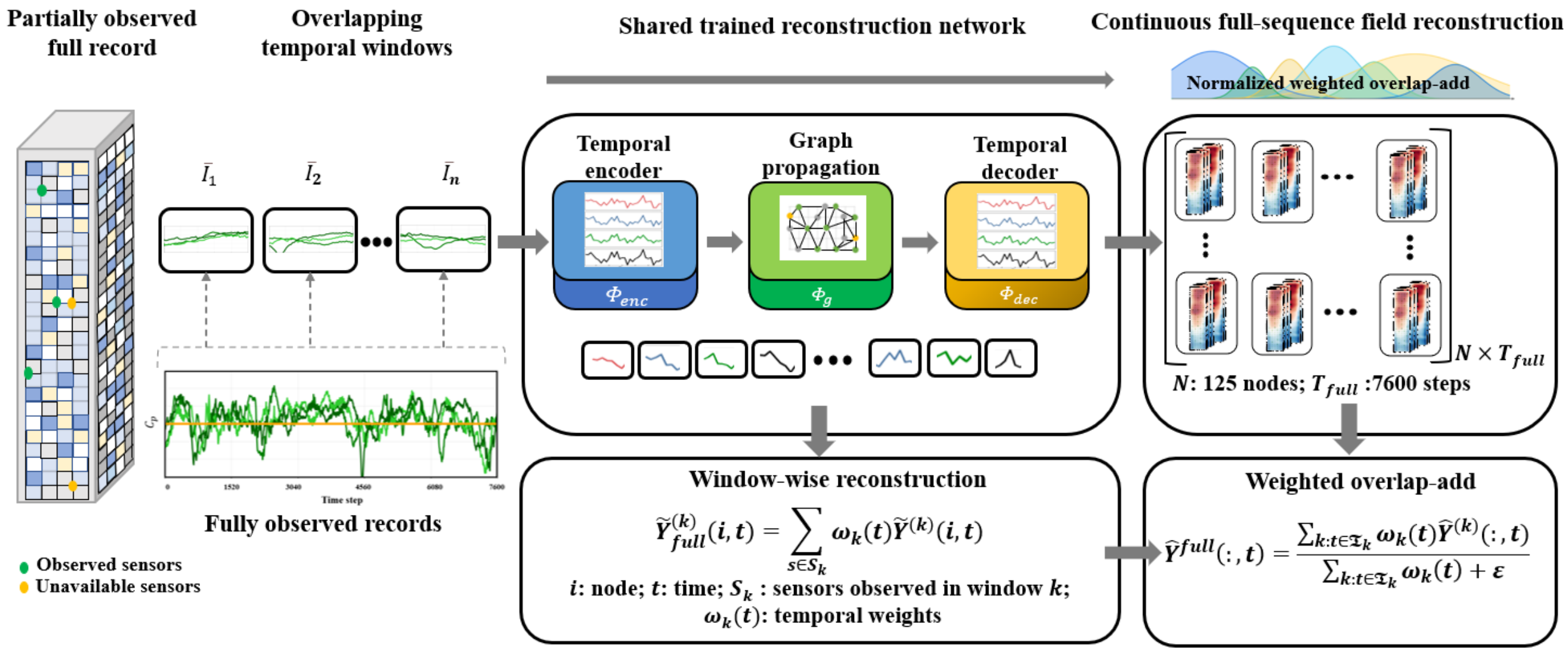

**Fig. 2** Full-sequence inference using overlapping window reconstruction and normalized weighted overlap-add aggregation.

Let $\mathbf{X}^{full} \in \mathbb{R}^{|\mathcal{V}|\times T_{full}}$ denote the partially observed pressure measurements over a full record of length $T_{full}$, together with the associated observation-availability indicator and conditioning variables. Because the trained model reconstructs the field over intervals of fixed duration, the full record is partitioned into a sequence of overlapping windows. For a window index $k$, let $\mathfrak{T}_k \subset \{1, \dots, T_{full}\}$ denote the temporal indices covered by that window, and let $\mathbf{X}^{(k)}$ be the corresponding partially observed segment. The trained reconstruction network is then applied independently to each segment to obtain a window-level full-field estimate $\widehat{\mathbf{Y}}^{(k)}$.

A direct concatenation of successive window predictions is generally not suitable because reconstruction quality often degrades near window boundaries, particularly when temporal encoding and decoding are performed over finite receptive fields. To mitigate this effect, the inference procedure employs overlapping windows together with weighted aggregation. Each window-level prediction is multiplied by a smooth temporal weighting function $\omega(\tau)$, defined over the local window coordinate $\tau$, such that lower weights are assigned near the interval edges and higher weights near the center. The final full-sequence reconstruction is then formed through normalized overlap-add accumulation:

$$\widehat{\mathbf{Y}}^{full}(:,t) = \frac{\sum_{k:t\in\mathfrak{T}_k} \omega_k(t)\widehat{\mathbf{Y}}^{(k)}(:,t)}{\sum_{k:t\in\mathfrak{T}_k} \omega_k(t)+\varepsilon}, \qquad t = 1, \dots, T_{full} \tag{10}$$

where $\omega_k(t)$ is the weight assigned by window $k$ to the global time index $t$, and $\varepsilon$ is a small positive constant introduced for numerical stability. This overlap-and-merge strategy reduces stitching artifacts and restores temporal continuity without requiring an additional smoothing post-processing stage.

The same observation-mask and conditioning logic used during training is retained during inference, although the role of masking differs from that used in the learning stage. During training, masking patterns are varied to expose the model to diverse partial-observation scenarios and improve robustness. During evaluation or deployment, the masking pattern is defined by the reconstruction scenario of interest, such as a prescribed sensor-outage configuration used for repeatable benchmarking. The observation-availability indicator is therefore propagated through the full-sequence inference pipeline exactly as in training, allowing the network to distinguish unavailable measurements from physically meaningful signal values throughout reconstruction.

In practical terms, full-sequence inference consists of four steps: defining the reconstruction scenario and its associated observation mask, partitioning the input record into overlapping windows, reconstructing each window independently with the trained network, and merging the resulting window-level predictions through normalized weighted overlap-add aggregation. The exact choices of window length, overlap ratio, and weighting function are implementation-specific and are therefore reported later in the case-study and implementation section.

## 3. Case Study and Results

### 3.1 Case study description and aerodynamic pressure dataset

The proposed framework is evaluated using wind-tunnel façade-pressure data from the Tokyo Polytechnic University (TPU) aerodynamic database. The case study considers a rigid high-rise building model tested in an open-circuit wind tunnel under a simulated atmospheric boundary layer, providing time-resolved pressure measurements suitable for assessing both masked-sensor recovery and full-field reconstruction under realistic aerodynamic variability. The model is instrumented with 500 pressure taps distributed over the four façades, corresponding to 125 taps per façade, arranged on a 25×5 grid.

The measured quantity is the pressure coefficient, $C_p$, sampled at 1000 Hz. The database includes 11 wind directions spanning 0° to 50° in 5° increments. Under the adopted convention, 0° corresponds to flow normal to the reference façade, while the remaining directions represent successive clockwise rotations of the incoming wind. This directional set provides a range of loading conditions associated with distinct spatial pressure distributions and temporal signatures, making it appropriate for evaluating reconstruction performance under varying aerodynamic regimes.

The analysis focuses on three façades: the windward façade, the right-side façade, and the leeward façade. This selection is intended to span the principal aerodynamic pressure regimes relevant to façade-pressure reconstruction. The windward face is primarily influenced by impingement-driven loading, whereas the lateral and leeward faces are more strongly affected by separated flow, shear-layer development, and wake-induced fluctuations, leading to richer spatial variability and a more challenging reconstruction setting. Only one lateral façade is included because, for the isolated prismatic building considered here, the two side faces are expected to exhibit analogous behavior under mirrored flow conditions; the right-side façade is therefore taken as representative of lateral-face reconstruction difficulty.

For each wind direction, a fixed-length segment of 7600 samples is retained from the available record. Using a common segment length across directions provides a consistent basis for training, holdout evaluation, and full-sequence reconstruction experiments. The reconstruction target is therefore the full façade field over the retained interval, reconstructed from sparse and partially missing measurements. The graph representation, sensor configuration, preprocessing procedure, and data-partitioning strategy adopted prior to training are described in the next subsection.

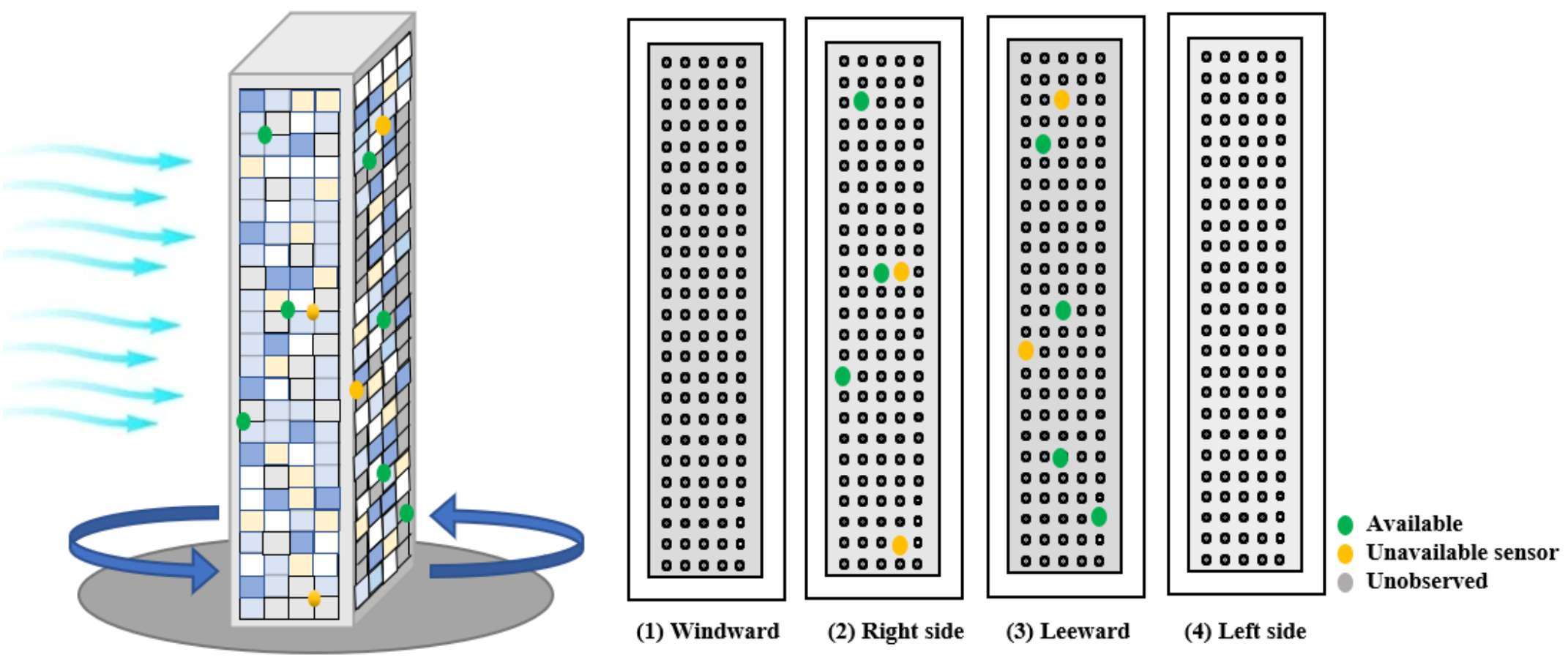

**Fig. 3** Experimental case-study setup: (a) wind-tunnel model and wind-direction convention; (b) façades pressure-tap grid (25×5, 125 taps per façade).

### 3.2 Spatial domain representation, graph construction, and sensor configuration

Each analyzed façade is represented as a discrete spatial domain consisting of 125 nodes arranged on the same 25×5 pressure-tap grid in Section 3.1. The façade nodes are indexed in a column-wise manner, so that each spatial location is assigned a unique global node identifier based on its row and column position. This indexing is used consistently for graph construction, sensor assignment, and reporting of reconstructed node locations.

The reconstruction model operates on a unified graph containing all 125 façade nodes, with instrumented locations treated as a subset of nodes in the same domain. The base graph connectivity is defined from the façade grid topology using nearest-neighbor links in the horizontal and vertical directions. In practice, each node is connected to its right neighbor and downward neighbor when these exist, and the resulting edges are stored as undirected connections through bidirectional edge pairs. Self-loops are not included in the base graph. This construction yields a regular grid graph that preserves the physical adjacency structure of the façade discretization.

To improve information transfer among instrumented locations, additional sensor-to-sensor edges are added to the base grid graph. These supplementary edges are defined directly in the global node indexing and follow three deterministic rules: adjacent sensors along the same row are connected in a chain, adjacent sensors along the middle column are connected vertically, and selected diagonal sensor pairs satisfying the implemented geometric conditions are connected when both endpoints belong to the instrumented set. The final edge set is obtained by merging the grid edges and the sensor-specific edges while removing duplicates. The resulting graph therefore preserves the global façade geometry while enriching connectivity pathways among the available sensors. The instrumented configuration comprises 24 observed nodes per façade, and the same sensor layout is used for the windward, right-side, and leeward façades to ensure consistency across training and evaluation. The selected sensors are distributed over the façade surface to provide coverage in both height and width, while the remaining 101 nodes are treated as non-instrumented locations to be reconstructed by the proposed framework.

For graph-feature construction, spatial coordinates are assigned to all 125 nodes using the façade grid indices and are combined with the observation-availability indicator and wind-direction encoding in the node-token representation introduced in Section 2.3. This allows graph message passing to remain conditioned on spatial position, measurement availability, and loading direction. Two indexing conventions are used in the remainder of the paper and should be distinguished clearly: global node indices in the 125-node façade graph, used for graph construction and reporting of unobserved-node results, and local sensor indices in the ordered 24-sensor list, used in masked-sensor plots and fixed masking definitions.

### 3.3 Data preprocessing

The preprocessing pipeline is applied separately to each wind direction after extraction of the retained directional records described in Section 3.1. For a given direction, the full façade pressure-coefficient field is arranged as a time-resolved array over the 125 façade nodes, using the retained record length of 7600 samples. A contiguous 80–20 temporal split is then applied prior to window sampling, yielding a training segment of 6080 samples and a holdout segment of 1520 samples per direction. This ordering is adopted to avoid temporal leakage between training and holdout data.

Normalization is performed independently for each wind direction and façade node using a z-score transformation based exclusively on statistics computed from the corresponding training segment. Let $C_p^{(d)}(n, t)$ denote the pressure coefficient at time index $t$, node $n$, and wind direction $d$. For each direction-node pair, the mean and standard deviation are computed from the training segment only, and both the training and holdout segments are normalized using these training statistics. This prevents leakage of holdout information into the normalization stage and ensures that all reported evaluations are carried out under a consistent preprocessing protocol.

Training samples are constructed as fixed-length temporal windows drawn after the train–holdout split has been established. In the present implementation, the window length is 200 samples. Rather than using deterministic sliding windows during training, windows are sampled randomly from the directional training pool, with the wind direction and temporal start index selected independently for each sampled sequence. This increases variability across mini-batches while preserving the integrity of the temporal partition. Each sampled window is also associated with the corresponding wind-direction conditioning feature used by the reconstruction network.

### 3.4 Model implementation and training settings

The case-study implementation follows the encoder–graph–decoder formulation introduced in Section 2 and uses a fixed configuration across the reported façade experiments. The model is trained for multi-direction reconstruction on the 125-node façade graph, with input windows sampled across wind-direction records and conditioned using the adopted wind-direction encoding. The temporal encoder is implemented as a dilated residual Conv1D network with four residual blocks (kernel size 3; dilations 1, 2, 4, and 8), followed by two strided Conv1D layers that reduce the temporal dimension from 200 to 50 while mapping the signal to a latent channel dimension of 128. Group normalization and GELU activation are used throughout the encoder. Spatial propagation is performed by a four-layer graph attention network with hidden dimension 128, four attention heads, residual connections, layer normalization, and ELU activation. For each latent-time slice, the node token is formed by concatenating the latent feature vector with five auxiliary attributes (two spatial coordinates, one observation-mask flag, and two wind-direction features) resulting in an input token dimension of 133 to the graph module. The decoder reconstructs the full temporal window in a single forward pass using an initial Conv1D projection, two ConvTranspose1D layers for temporal upsampling from the latent grid to the original window length, and three residual temporal convolutional blocks with GroupNorm and GELU activations.

Optimization is performed for 300 epochs with 10 optimization steps per epoch and a mini-batch size of 8 windows. AdamW is used with a learning rate of $10^{-4}$ and zero weight decay, with gradient clipping at a maximum norm of 1.0. The reconstruction loss follows the composite formulation introduced in Section 2.4, combining a Huber amplitude term, a temporal-difference term, and a one-sided FFT spectral-magnitude term. In the implemented configuration, the Huber parameter is $\beta = 1.0$, the temporal-difference weight is $\lambda_d = 0.3$, and the spectral weight is $\lambda_f = 0.05$. The subset-weighted reconstruction loss uses $\omega_m = 1.5$, $\omega_o = 0.1$, and $\omega_u = 1.0$ for masked instrumented nodes, available instrumented nodes, and non-instrumented nodes, respectively, while the latent consistency term is weighted by $\lambda_z = 0.2$. The latent consistency branch is implemented in teacher–student form: the reference branch computes the latent representation from the unmasked input, whereas the main reconstruction branch operates on the masked input and graph-conditioned latent state, with consistency enforced only at masked sensor locations. A curriculum masking strategy is adopted during training, with one instrumented sensor masked per sampled window in the initial stage and two masked thereafter, to progressively increase reconstruction difficulty under partial observation.

### 3.5 Quantitative results and aggregated performance

Quantitative reconstruction performance is reported for two targets reflecting the intended application of the framework in façade-pressure monitoring. The first target addresses robustness to sensor outage and is assessed by withholding measurements from a subset of instrumented sensors at inference time (2 out of 24 sensors) and comparing reconstructed signals against their measured references. The second target addresses full-field completion and is assessed by reconstructing pressure signals at non-instrumented façade nodes, which are never provided as input, and comparing them against corresponding measured references available in the dataset. All statistics reported in this subsection are computed on the holdout temporal segment of each wind-direction record under the evaluation protocol described in Sections 3.3–3.4. Performance is quantified using both time-domain and frequency-domain measures. Time-domain accuracy is evaluated using RMSE and MAE, and temporal agreement is summarized using the Pearson correlation coefficient. Dynamic fidelity is assessed in the frequency domain by computing the power spectral density (PSD) via Welch's method and quantifying band-limited discrepancies through the relative error in PSD-integrated energy (bandpower) over predefined frequency ranges. Results are reported separately for masked instrumented sensors and non-instrumented nodes, and are aggregated across wind directions for each façade as mean ± standard deviation.

Table 1 summarizes aggregated reconstruction performance for masked instrumented sensors under the sensor-outage scenario. Across all wind directions, the mean masked-sensor RMSE is 0.062 (±0.028), with MAE 0.045 (±0.019), correlation 0.942, and PSD discrepancy 0.133.

**Table 1**. Aggregated reconstruction performance for masked instrumented sensors under the sensor-outage scenario, reported over the holdout interval as mean ± standard deviation across wind directions for each façade.

| Masked locations | Facade | RMSE | MAE | Correlation | PSD (%) |
|---|---|---|---|---|---|
| | Windward | 0.041(±0.008) | 0.032 (±0.007) | 0.98 | 0.098 |
| | Right-side | 0.09 (±0.053) | 0.066 (±0.038) | 0.92 | 0.20 |
| | Leeward | 0.054 (±0.022) | 0.037(±0.013) | 0.925 | 0.10 |

Table 2 reports the corresponding aggregated performance for non-instrumented façade nodes under the full-field completion scenario. Across all wind directions, the mean full-field RMSE is 0.077 (±0.027), with MAE 0.057 (±0.019), correlation 0.908 (±0.037), and PSD discrepancy 0.153 (±0.034). The façade-wise results reported in Tables 1 and 2 show consistent trends, with lower reconstruction errors on the windward façade and higher errors on the right-side and leeward façades. This behavior is physically consistent with the increased spatio-temporal complexity associated with separated flow and wake effects on side and leeward surfaces.

Figure 4 provides a complementary direction-wise view of reconstruction accuracy, reporting RMSE as a function of wind direction for both masked instrumented sensors and non-instrumented nodes across the three façades. The figure highlights the same trends observed in the aggregated statistics, while also revealing directional variability, particularly on the right-side façade where errors are consistently higher across most wind directions.

**Table 2**. Aggregated reconstruction performance for non-instrumented façade nodes under the full-field completion scenario, reported over the holdout interval as mean ± standard deviation across wind directions for each façade.

| Masked/ unobserved locations | Facade | RMSE | MAE | Correlation | PSD (%) |
|---|---|---|---|---|---|
| | Windward | 0.055 (±0.011) | 0.04 (±0.008) | 0.94 | 0.14 |
| | Right-side | 0.12 (±0.053) | 0.09 (±0.038) | 0.88 | 0.22 |
| | Leeward | 0.056 (±0.018) | 0.04 (±0.012) | 0.905 | 0.10 |

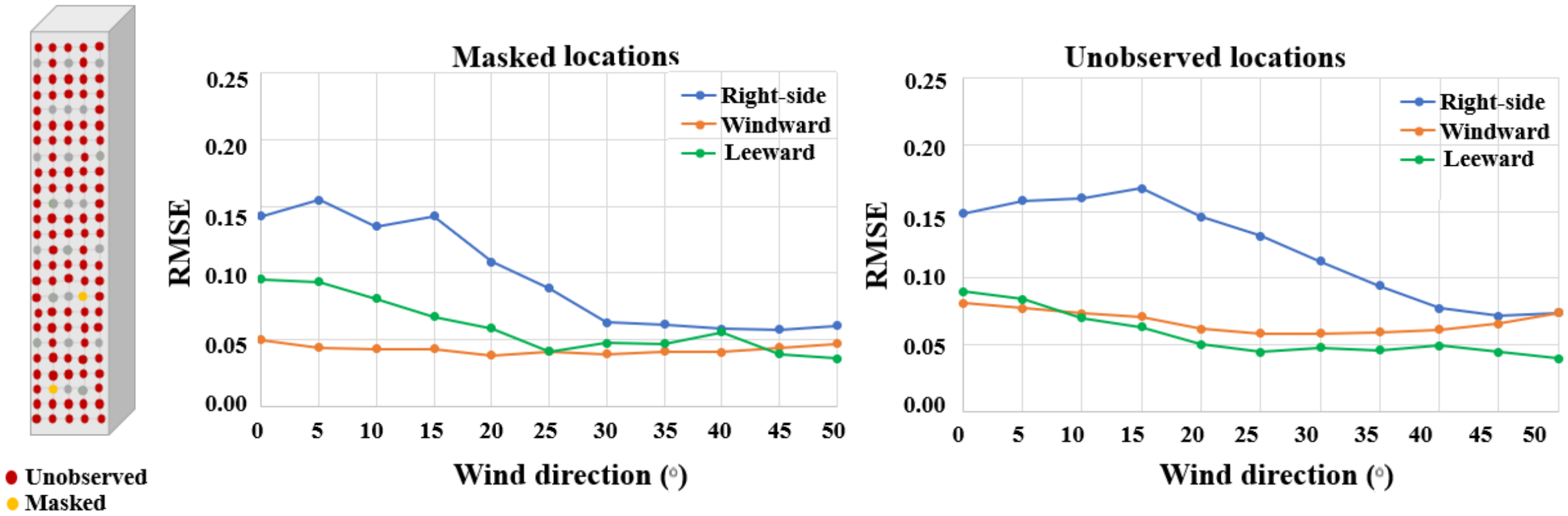

**Fig. 4** Direction-wise RMSE for masked instrumented sensors and non-instrumented nodes on the windward, right-side, and leeward façades.

To assess robustness to the spatial configuration of sensor outages, Table 3 reports reconstruction performance for masked instrumented sensors aggregated over multiple sensor-outage realizations, each corresponding to a distinct outage pattern. The results show that reconstruction performance remains consistently stable on the windward façade, with low variability across realizations and high correlation values (≈0.97–0.99). In contrast, the right-side façade exhibits both higher reconstruction errors and greater variability, indicating increased sensitivity to the location of missing sensors. The leeward façade demonstrates intermediate behavior, with moderate variability and consistently lower errors than the right-side façade. Overall, these results confirm that while the framework is robust to sensor outages on the windward façade, its sensitivity increases in more complex flow regions, particularly on the right-side façade where separated and wake-driven dynamics dominate.

**Table 3**. Reconstruction performance for masked instrumented sensors under multiple sensor-outage realizations, reported over the holdout interval as mean ± standard deviation across realizations for each façade.

| Masked locations | Facade | RMSE | MAE | Correlation | PSD (%) |
|---|---|---|---|---|---|
| | Windward | 0.048 (±0.007) | 0.040 (±0.008) | 0.985 | 0.09 |
| | Right-side | 0.115 (±0.046) | 0.089 (±0.031) | 0.930 | 0.18 |
| | Leeward | 0.045 (±0.018) | 0.032 (±0.012) | 0.941 | 0.07 |
| | Windward | 0.063 (±0.006) | 0.051 (±0.007) | 0.971 | 0.08 |
| | Right-side | 0.128 (±0.06) | 0.097 (±0.044) | 0.890 | 0.27 |
| | Leeward | 0.056 (±0.022) | 0.038 (±0.015) | 0.943 | 0.10 |
| | Windward | 0.043 (±0.006) | 0.034 (±0.006) | 0.980 | 0.04 |
| | Right-side | 0.128 (±0.05) | 0.098 (±0.039) | 0.923 | 0.26 |
| | Leeward | 0.051 (±0.023) | 0.034 (±0.015) | 0.950 | 0.09 |
| | Windward | 0.046 (±0.004) | 0.037 (±0.005) | 0.978 | 0.11 |
| | Right-side | 0.116 (±0.048) | 0.088 (±0.032) | 0.920 | 0.17 |
| | Leeward | 0.050 (±0.015) | 0.035 (±0.011) | 0.945 | 0.08 |

### 3.6 Representative façade-wise case studies

The aggregated metrics reported in Section 3.5 establish the overall reconstruction performance across façades and wind directions. The present subsection complements those results with representative façade-wise case studies showing time-domain agreement, spectral consistency, and instantaneous spatial reconstruction under the same evaluation protocol. The examples are organized by façade (i.e., windward, right-side, and leeward) to reflect progressively more complex aerodynamic regimes and to provide physical context for the façade-wise trends observed in the aggregated results.

For the windward façade, Fig. 5 illustrates a representative masked-sensor reconstruction case at $\theta = 25°$. The reconstructed responses at the two masked instrumented sensors closely follow the measured time histories over the holdout interval, with RMSE/MAE values of 0.015/0.012 and 0.019/0.015, and corresponding correlations of 0.997 and 0.995. The PSD comparison further indicates good preservation of the dominant energy-containing content. These results are consistent with the lower aggregate errors reported for the windward façade in Section 3.5 and suggest that the remaining discrepancies are mainly associated with short-duration fluctuations rather than systematic bias or temporal drift.

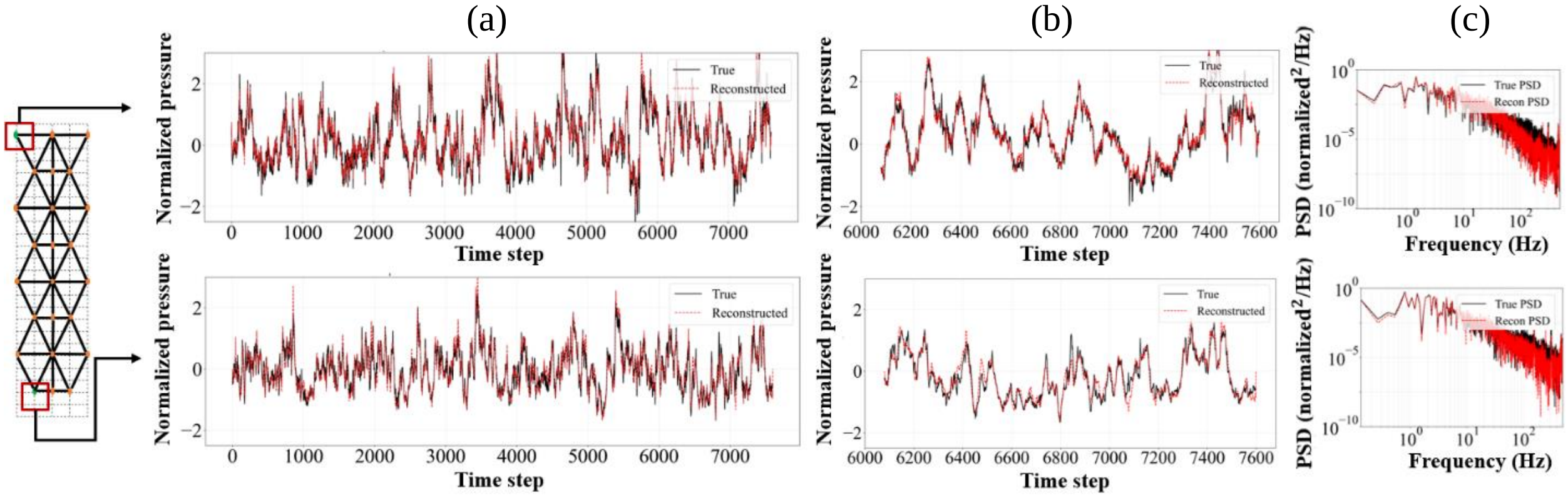

**Fig. 5** Windward façade case study at $\theta = 25°$ : measured and reconstructed pressure-coefficient time histories at two masked sensors over (a) the full record and (b) the holdout interval, with (c) the corresponding PSD comparison.

Field-level reconstruction behavior is further examined in Fig. 6 through instantaneous spatial snapshots. The reference and reconstructed pressure fields exhibit consistent large-scale spatial organization at the selected instants, including the location and extent of the dominant suction regions. The associated snapshot-level spatial RMSE values are 0.0516 and 0.0514. Together, these results show that the model preserves coherent full-field structure on the windward façade beyond agreement at the masked sensor locations.

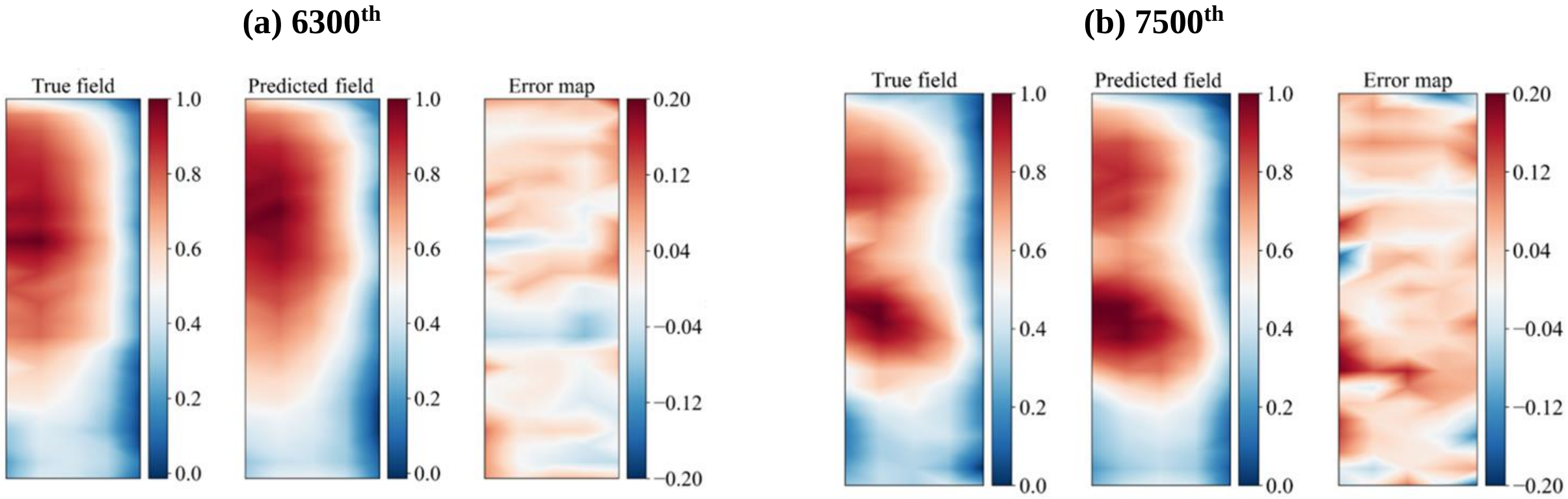

**Fig. 6** Windward façade spatial reconstruction at two representative instants: reference field, reconstructed field, and error map.

For the right-side façade, reconstruction difficulty increases, consistent with the higher aggregate errors reported in Section 3.5. Figure 7 presents a representative lateral-façade case at $\theta = 5°$, showing that the reconstructed masked-sensor time histories retain the dominant temporal structure while exhibiting larger localized deviations during intermittent higher-frequency fluctuations. Over the holdout interval, the RMSE/MAE values at the two masked sensors are 0.157/0.116 and 0.082/0.064, with correlations of 0.877 and 0.9833. The PSD comparison indicates that the principal energy-containing frequencies remain captured, while differences increase in the higher-frequency range, consistent with the broader-band dynamics associated with side-face loading.

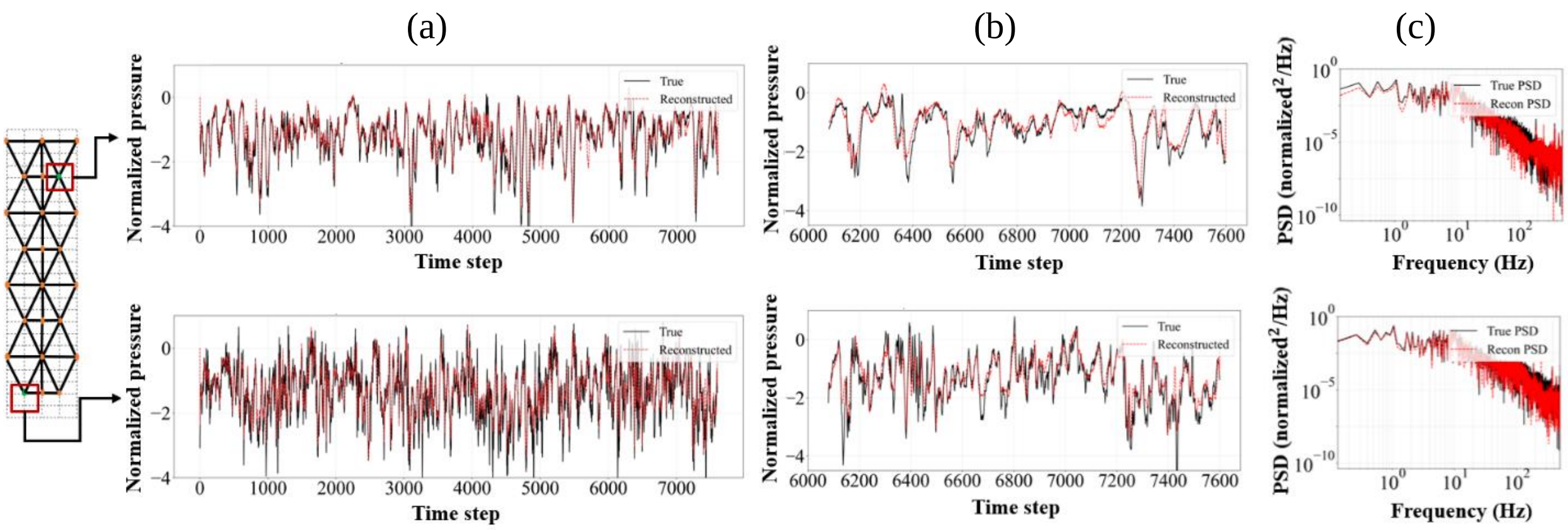

**Fig. 7** Right-side façade case study at $\theta = 5°$: measured and reconstructed pressure-coefficient time histories at two masked sensors over (a) the full record and (b) the holdout interval, with (c) the corresponding PSD comparison.

Figure 8 provides the corresponding spatial validation. The reconstructed fields recover the main spatial gradients and suction regions observed in the reference snapshots, while the error maps highlight localized deviations in regions associated with strong suction peaks and shear layer gradients. Snapshot spatial RMSE values of 0.131 and 0.127 quantify the field-level agreement and provide a direct visual explanation for the increased error levels relative to the windward case.

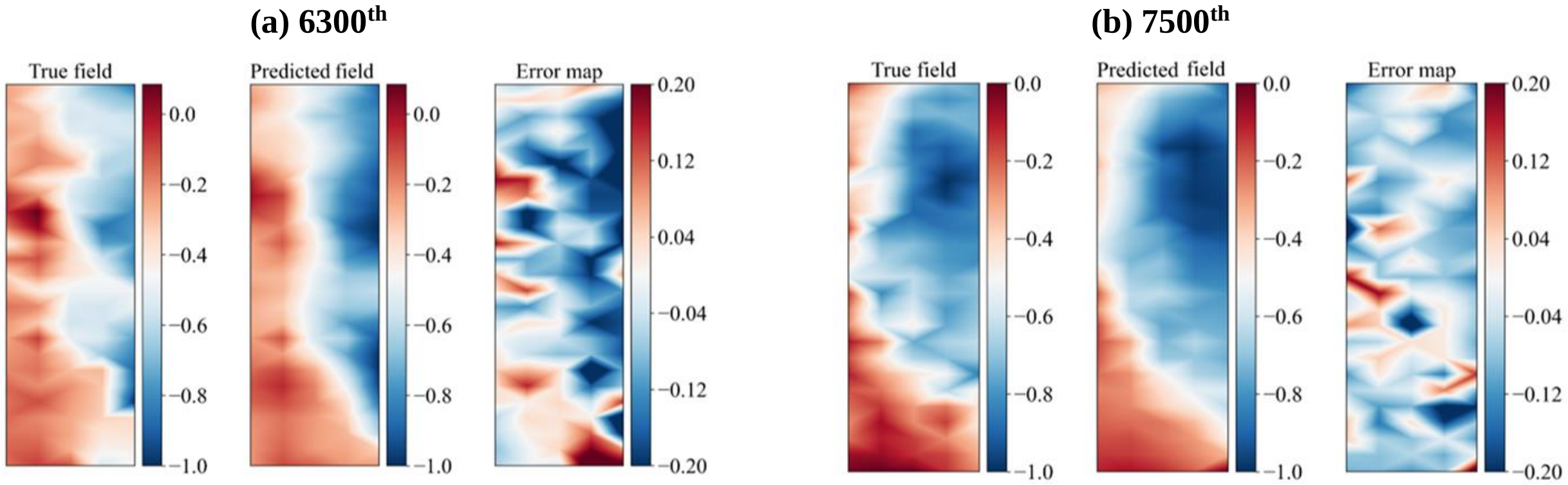

**Fig. 8** Right-side façade spatial reconstruction at two representative instants: reference field, reconstructed field, and error map.

Figure 9 shows a representative leeward case at $\theta = 50°$, where the reconstructed masked-sensor signals maintain the principal temporal organization but display larger localized differences during sharp events and higher-frequency fluctuations. The corresponding holdout RMSE/MAE values at the two masked sensors are 0.03/0.023 and 0.016/0.012, with correlations of 0.956 and 0.993. The frequency-domain comparison indicates that dominant low-frequency energy is retained, whereas the largest deviations occur in the higher-frequency content, consistent with the leeward aggregate performance reported in Section 3.5.

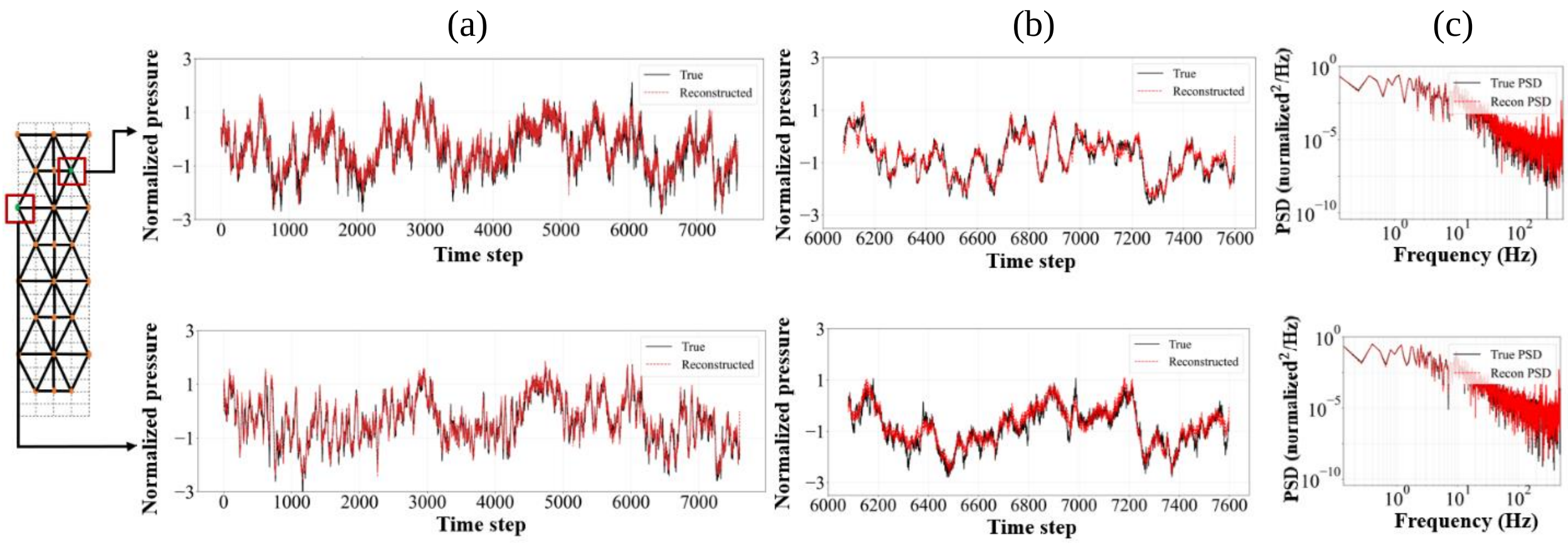

**Fig. 9** Leeward façade case study at $\theta = 50°$: measured and reconstructed pressure-coefficient time histories at two masked sensors over (a) the full record and (b) the holdout interval, with (c) the corresponding PSD comparison.

Figure 10 presents the associated spatial snapshots and error maps. The reconstruction captures the overall wake-driven pressure-field organization at the reported instants, while discrepancies become more pronounced in high-gradient regions associated with wake shear layers and flow separation zones, due to the under-resolution of small-scale turbulent structures under sparse sensing conditions. Snapshot spatial RMSE values of 0.041 and 0.071 quantify the field-level agreement and provide a direct link between the leeward flow physics and the elevated reconstruction errors observed in the aggregated results.

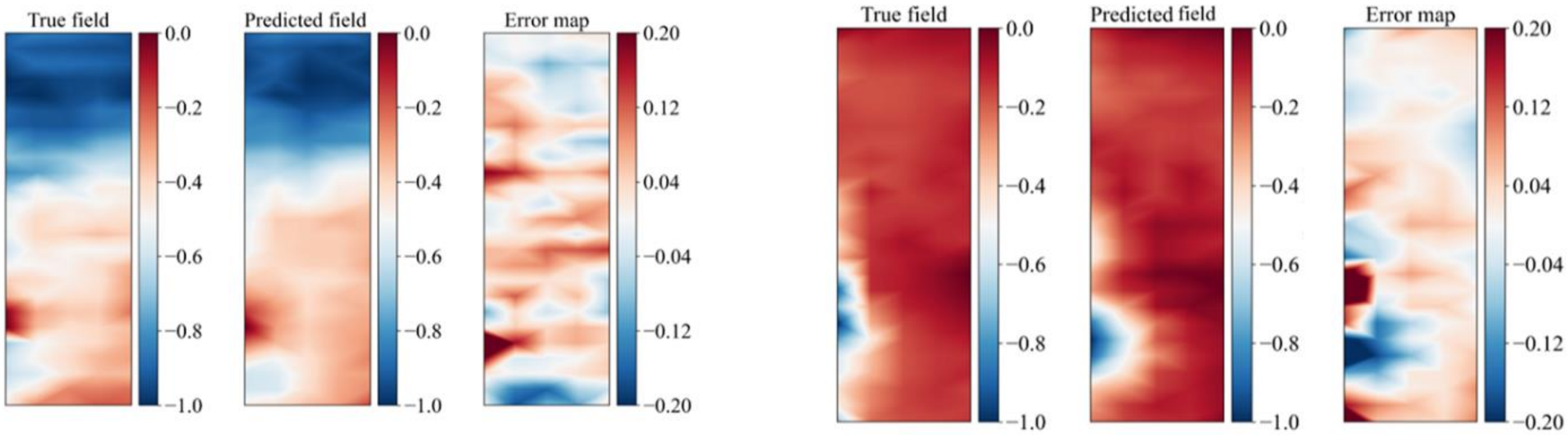

**Fig. 10** Leeward façade spatial reconstruction at two representative instants: reference field, reconstructed field, and error map.

Across all three façades, the representative case studies confirm that the framework preserves the dominant temporal evolution and principal spectral characteristics at masked sensors while producing spatially coherent full-field reconstructions. The progressive increase in reconstruction difficulty from windward to right-side and leeward façades is reflected consistently in both the aggregated metrics of Section 3.5 and the representative time-domain, spectral, and spatial validations presented here, indicating that the observed error hierarchy is primarily governed by façade-dependent aerodynamic complexity.

## 4. Discussion

### 4.1 Discussion of main findings

The results show that the proposed framework can reconstruct time-resolved façade pressure fields from sparse and partially missing measurements within a single unified formulation. By combining temporal feature extraction with graph-based spatial propagation and explicit observation masking, the method addresses both masked-sensor recovery and reconstruction at non-instrumented locations over the same time interval, rather than treating these tasks as separate stages. This unified design is important in the present setting because sparse sensing and temporary sensor outage are not exceptional conditions, but inherent features of façade-pressure monitoring.

A consistent performance hierarchy is observed across façades, with the best reconstruction generally obtained on the windward façade and lower accuracy on the lateral and leeward façades. This trend is physically meaningful. Windward loading is dominated by comparatively smoother impingement-driven pressure organization, whereas the lateral and leeward façades are influenced more strongly by separation, shear-layer dynamics, and wake effects, which introduce broader-band fluctuations, sharper localized events, and greater spatio-temporal variability. The results, therefore, indicate that reconstruction difficulty is governed not only by sensor sparsity but also by the underlying aerodynamic complexity of the target field.

The time-history, spectral, and spatial validations further clarify the character of the reconstruction errors. In all three façades, the dominant temporal evolution and principal spectral content are reproduced well, and the reconstructed spatial snapshots retain the main pressure-field organization. The remaining discrepancies tend to concentrate in localized high-frequency or intermittent components, especially on the lateral and leeward façades. This suggests that the framework is already effective for recovering the large-scale and dynamically dominant structure of the façade pressure field, while the most difficult aspect remains the accurate reconstruction of localized, rapidly varying fluctuations associated with separated and wake-dominated flow regions.

From a practical standpoint, these findings support the feasibility of outage-tolerant façade-pressure field completion from sparse instrumentation using a single end-to-end model. The fact that the same framework can recover temporarily unavailable instrumented signals while simultaneously reconstructing non-instrumented locations is particularly relevant for monitoring-oriented applications, where sensing density is limited and measurement availability may vary over time. In that sense, the present results establish not only predictive accuracy but also the operational relevance of sensor-outage-aware spatio-temporal graph reconstruction for aerodynamic pressure monitoring.

### 4.2 Two-stage predictive extension

The reconstruction framework developed in this study estimates a spatially complete façade pressure field over a given time interval from partially observed measurements over that same interval. For operational monitoring, however, an additional capability may be desirable: estimating façade pressures at future times. A natural extension is therefore a two-stage predictive formulation in which future pressure signals are first forecast at the available instrumented sensors and then mapped to a full-field estimate through the reconstruction model. In this sense, the present framework can serve as the second stage of a modular predictive pipeline rather than being modified into a direct full-field forecasting model.

In the first stage, a forecasting model $g_{\psi}$ operates only on the instrumented sensor channels and predicts their future pressure time series over a horizon $H$ from a recent history window. Because this forecasting task is restricted to the sensor set, it remains much lower-dimensional than direct future full-field prediction and can be implemented using standard temporal forecasting architectures. In the second stage, the proposed reconstruction model $f_{\theta}$ is used as a spatial completion operator. The forecasted sensor signals are placed on the façade graph at the instrumented nodes; non-instrumented nodes remain unobserved, and the same mask-aware inference pathway is applied to generate a spatially complete future field estimate $\widehat{\mathbf{Y}}^{fut}$ over the forecast horizon. This formulation isolates temporal prediction to the sensor space while leveraging the learned spatial correlations and façade connectivity captured by the reconstruction model.

Figure 11 provides a proof-of-concept demonstration of this two-stage pipeline for the windward façade at $\theta = 20°$and forecasting horizon $H = 1.52$ s. The sensor-level forecasting performance over the horizon is summarized by RMSE = 0.004, MAE = 0.003, and correlation = 0.99. The corresponding full-field prediction is then examined in two forms: a reference-input case, in which the reconstruction model is driven by measured future sensor signals, and a two-stage predictive case, in which the forecasted sensor signals drive the same reconstruction model. For the reference-input case, the full-field completion accuracy at non-instrumented nodes is RMSE = 0.044, MAE = 0.035, and correlation = 0.91. In the two-stage predictive case, the corresponding full-field metrics are RMSE = 0.051, MAE = 0.04, and correlation = 0.905. The difference between these two outputs provides a direct measure of error propagation from sensor forecasting to full-field estimation; for this demonstration case, the increase in non-instrumented-node RMSE from the reference-input case to the two-stage case is 0.007, indicating limited error amplification through the forecasting–reconstruction pipeline.

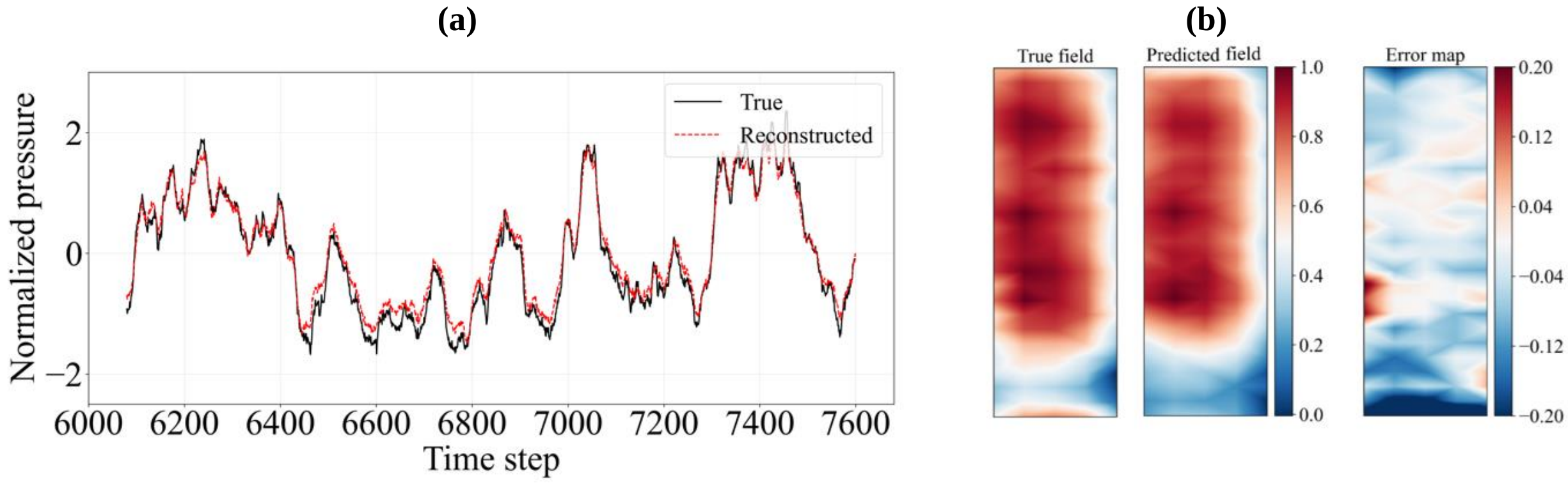

**Fig. 11** Two-stage predictive extension on the windward façade ($\theta = 20°$, horizon $H$ =1.52 s): (a) sensor-level forecasting at instrumented locations and (b) resulting full-field prediction from measured versus forecasted sensor inputs at 7480$^{th}$ instant.

This two-stage formulation preserves the reconstruction model as a dedicated field-completion operator while enabling predictive use through a modular sensor-forecasting front end. It also makes explicit that end-to-end predictive accuracy depends on both the quality of the sensor-level forecasts and the sensitivity of the reconstruction mapping to perturbations in the sensor inputs. As a result, the present demonstration should be interpreted as a promising extension of the reconstruction framework and a practical pathway toward predictive façade-pressure monitoring, rather than as a fully developed predictive contribution on the same footing as the core reconstruction results. More comprehensive assessment across façades, wind directions, and forecast horizons remains a natural direction for future work.

### 4.3 Limitations and practical considerations

Several limitations should be considered when interpreting the present results and assessing practical deployment. First, the evaluation is conducted for a single high-rise building configuration using a single experimental database. Although the selected case study spans multiple wind directions and three façade regimes of increasing aerodynamic complexity, transferability to other building geometries, surrounding exposure conditions, and flow environments remains to be demonstrated. Extending the framework to such settings would require either additional multi-configuration training data or an explicit transfer-learning strategy.

Second, reconstruction performance depends not only on the learning architecture but also on the sensor layout and on the spatial location of missing measurements. While the framework is evaluated under multiple outage realizations, sensitivity may increase when unavailable sensors lie in regions that are particularly informative for the pressure field, such as edges, corners, or zones of strong suction. This consideration is especially relevant for monitoring-oriented applications, where sensor loss is not controllable and where practical sensor placement should ideally be designed jointly with reconstruction robustness in mind.

Third, the results indicate that the proposed framework reconstructs the dominant temporal behavior and principal spectral content reliably, whereas the largest discrepancies tend to occur in localized high-frequency or intermittent components, especially on the lateral and leeward façades. The practical significance of this behavior depends on the intended downstream use. Applications primarily concerned with large-scale loading trends or dominant dynamic content may tolerate such residual smoothing, whereas applications requiring accurate reproduction of short-duration local peaks may require additional modeling emphasis on extremes, increased sensing density in critical regions, or alternative loss formulations tailored to localized transient behavior.

Finally, the two-stage predictive extension highlights a separate practical limitation: end-to-end predictive field accuracy is constrained by error propagation from sensor-level forecasting to spatial field completion. Even when the reconstruction model itself is stable, uncertainty in the forecasted sensor signals will propagate to the full-field estimate, and this sensitivity is expected to become more pronounced in separation- and wake-dominated regimes. For that reason, predictive deployment should assess both stage-wise and end-to-end performance and examine how sensor-level forecast uncertainty translates into uncertainty in the reconstructed future field across façades, wind directions, and forecast horizons.

## 5. Conclusion

This study presented a sensor-outage-aware spatio-temporal graph reconstruction framework for completing high-rise façade pressure fields from sparse measurements under constrained sensing and sensor-outage conditions. The proposed method combines temporal feature extraction with graph-based spatial propagation on a unified façade-domain representation, allowing the reconstruction of both temporarily unavailable instrumented sensors and non-instrumented façade locations over the same time interval as the available measurements. This formulation treats sparse sensing and missing observations as intrinsic aspects of the reconstruction problem rather than as separate post-processing issues. The case study based on wind-tunnel pressure coefficient data for a high-rise building showed that the framework provides reliable time-resolved reconstruction across windward, right-side, and leeward façades under multiple wind directions. The results demonstrated robust masked-sensor recovery and accurate full-field completion from sparse instrumentation, with a consistent façade-dependent performance hierarchy: reconstruction was generally most accurate on the windward façade and more challenging on the lateral and leeward façades, where separation and wake-driven dynamics introduce greater spatio-temporal complexity. Time-domain, spectral, and spatial validations further showed that the method preserves the dominant temporal evolution, principal dynamic content, and coherent large-scale pressure-field organization, while the largest residual discrepancies tend to remain localized in higher-frequency or intermittent components. Beyond reconstruction over a known interval, a two-stage predictive extension was also outlined, in which future sensor signals are first forecast at available instrumented locations and then mapped to future full-field pressure estimates through the proposed reconstruction model. This extension should be viewed as a promising operational pathway, since end-to-end predictive accuracy depends on both sensor-level forecasting performance and the sensitivity of the reconstruction mapping to forecast errors.

**Acknowledgment:** This work was supported by the Natural Sciences and Engineering Research Council of Canada (NSERC) [grant number CRSNG RGPIN 2022-03492].